\documentclass[aps,prl,showpacs,twocolumn]{revtex4-1}
\usepackage{amsmath,amssymb,latexsym}
\usepackage{fullpage,float,graphicx,mathrsfs,subfig}

\begin{document}

\title{Discrete Breathers at the Interface \\ between a Diatomic and a Monoatomic 
Granular Chain}
\author{C. Hoogeboom$^1$, G. Theocharis$^2$, P.G. Kevrekidis$^1$}
\affiliation{$^1$ Department of Mathematics and Statistics, University of Massachusetts, Amherst MA 01003-4515, USA  \\
$^2$ Graduate Aeronautical Laboratories (GALCIT),  \\ 
California Institute of Technology, Pasadena, CA 91125, USA}

\date{\today}

\begin{abstract}
	In the present work, we develop a systematic examination of the existence, stability and dynamical properties of a discrete breather at the interface between a diatomic and a monoatomic granular chain.
    We remarkably find that such an ``interface breather'' is {\it more robust} than its bulk diatomic counterpart throughout the gap of the linear spectrum.
	The latter linear spectral gap needs to exist for the breather state to arise and the relevant spectral conditions are discussed. 
	We illustrate the minimal excitation conditions under which such  an interface breather can be ``nucleated'' and analyze its apparently weak interaction with regular highly nonlinear solitary waveforms.
\end{abstract}

\maketitle

\section{Introduction}\label{s:intro}
	Over the past decade, granular chains have attracted significant attention \cite{nesterenko01,sen2007,vakakis2008}.
    This is because of the unusual nature that allows the tailoring of the response of the chain to be at the essentially linear, weakly nonlinear or strongly nonlinear regime \cite{vakakis2008}. 
	In particular, these chains of solid particles in the absence of static precompression are subject to purely nonlinear elastic contact interactions between their constituents leading to fundamentally novel features such as a vanishing sound speed (the so-called ``sonic vacuum'') \cite{nesterenko01}.
 	This renders the  linear and weakly nonlinear continuum approaches based on  the Korteveg-de Vries (KdV) equation \cite{KdV} invalid and places granular chains in a strongly nonlinear regime of wave dynamics. 
	Such a granular arrangement supports a new type of nearly compact highly tunable solitary waves that have been experimentally  and numerically observed in one-dimensional ($1D$) Hertzian granular systems \cite{Lazaridi1985,Sen1998,Sinkovits1995,coste97,Coste1999,Job2005,Vergara2005,Daraio2,Daraio3,Rosas2004,Rosas2007,Sokolow2007,pikovsky}.
 	While spherical bead chains have been mostly studied, similar responses can be obtained from cylindrical and elliptical particles, and fibrous layers \cite{Lambert1984}, among others. 

	More recently, it has been recognized that heterogeneous type settings bear significant additional potential for interesting nonlinear excitations.
 	In particular, in diatomic, triatomic etc. chains, the possibility of appropriately modified, nearly compact traveling waves was revealed in \cite{porter2008}, and \cite{porter2009}; on the other hand, more recently, the ability of such diatomic chains of spheres with Hertzian interactions under precompression to sustain discrete breather (DB) type excitations was proposed and experimentally corroborated.
    The theoretical \cite{Theo2010} and experimental realization \cite{boechler2009} of such nonlinear vibrational structures followed up on earlier works that had argued the emergence of nonlinear, defect-induced vibrational modes \cite{Theo2009}, even in the absence of precompression \cite{job}. 
    In such  acoustic vacuum settings, breathers have been argued to be metastable \cite{sen_prev}; it should also be mentioned that the linear dynamics of such diatomic crystals has been explored in the case of welded sphere contacts in~\cite{hladky}.

	It is at the latter junction of granular chains and nonlinear localized vibrations that the present work is centered.
 	In particular, we were motivated by the recent experimental investigation of \cite{khatri2009} examining the propagation of excitations at the interface between a monoatomic nonlinear chain and a linear medium to consider the possibility of formation of vibrational excitations at an interface.
  	In Refs. \cite{Theo2010}, and\cite{boechler2009}, we identified such modes in the ``bulk'' of the diatomic chain.
  	Specifically, we found two classes of solutions: (1) an unstable discrete breather that is centered on a heavy particle and characterized by a symmetric energy profile and (2) a potentially stable (at least in the vicinity of the linear spectrum's optical band edge) breather that is centered on a light particle and is characterized by an asymmetric energy profile.
    In the present work, we focus on the latter, more "robust" solution, and we compare it with its counterpart at the interface.
	It should be noted that such comparisons have recently been a focal point of interest in other fields, including e.g. nonlinear optics where such modes, the so-called nonlinear Tamm states, were  proposed theoretically \cite{makris2005,kartashov2006}, but also observed experimentally \cite{suntsov2006,rosberg2006,chen2007} in both one- and higher-dimensional settings.                                                         

	More specifically, we investigate the existence, stability, and dynamics of discrete breathers in a compressed  granular chain consisting of a diatomic part and a monoatomic part (cf. also with the experimental
context of Ref. \cite{nesterenko2005} where an interface between two monoatomic
chains was considered). 
	We are interested in such localized excitations at the interface between the two parts of the chain. 
	We first present the theoretical model, then analyze its linear spectrum;	using the latter as a starting point, and performing a continuation of the frequency parameter within the linear spectral gap, we are able to identify a family of nonlinear states localized at the interface. 
	We present the stability of these states and  compare/contrast it with that of the bulk structures in a diatomic chain identified in \cite{Theo2010}, and\cite{boechler2009}. 
	We remarkably find that these interface breathers are {\it more robust} than their diatomic bulk analogs.
	Next, we present another family of interface breathers that are shifted away from the interface by two lattice sites (a unit cell in the diatomic chain), and cease to exist as their frequency approaches the optical band edge.
	Lastly, we study the dynamics of this model, investigating the minimum requirements for generating these breathers in experimentally relevant settings, as well as examining the role of the interface in interactions with highly nonlinear, traveling solitary waves that are also supported by the model.

\section{Theoretical Setup}\label{s:model}
\begin{figure}
\includegraphics[width=.4\textwidth]{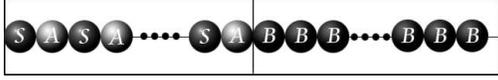}
\caption{
	The schematic set-up for the bead configuration: 
	steel - aluminum - . . . steel - aluminum - brass - brass - . . . - brass
}\label{f:setup}
\end{figure}
	We will focus our considerations on a one-dimensional array of beads, uniaxially compressed by a force $F_0$, which is modeled by zero-tension nonlinear spring-mass system.
	The presence of a precompressive force $F_0$, ensures the existence of an underlying linear model and hence a linear spectrum. 
	Note that the chain and the precompressive force will be assumed to be horizontal, so that the effects of gravity will be absent. 
	The equation of motion for the displacement $u_n$ of the $n$th bead (away from the chain boundaries) will be given by \cite{nesterenko01}
\begin{eqnarray}
m_n\ddot{u}_n &=&A_{n-1,n}[\Delta_{n-1,n}-(u_n-u_{n-1})]_+^{p}
\nonumber 
\\
&-& A_{n,n+1}[\Delta_{n,n+1}-(u_{n+1}-u_n)]_+^{p},\label{e:motion}
\end{eqnarray}
where $A_{n,n+1}$ is the elastic coefficient between beads $n$ and $n+1$, and depends on the exponent $p$ and the geometric and elastic properties of the beads $n$ and $n+1$. 
	In the Hertzian setting, the exponent takes the value $p=3/2$ and the coefficient $A_{n,n+1}$ is given by
\begin{equation}
A_{n,n+1}=\frac{4E_{n}E_{n+1}\sqrt{\dfrac{r_nr_{n+1}}{r_n+r_{n+1}}}}{3E_{n+1}(1-\nu_{n}^2)+3E_n(1-\nu_{n+1}^2)}\label{e:forcecoeff}
\end{equation}
where $r_n$, $E_n$, and $\nu_n$ denote, respectively, the radius, elastic (Young's) modulus, and Poisson ratio of the $n$th bead.
	$\Delta_{n,n+1}=(F_0/A_{n,n+1})^{2/3}$ is the amount of static overlap of the beads $n$ and $n+1$ (due to the precompression) under equilibrium conditions. 
	The bracket $[s]_+$ of Eq.~(\ref{e:motion}) takes the value $s$ if $s > 0$ and the value $0$ if $s \leq 0$. 
	This accounts for the fact that if two beads are not in contact, they do not exert any force on each other (zero-tension).
	Each bead (away from the boundary) has a local energy density associated with it \cite{fraternali2008} given by
\begin{equation}
e_n=\frac{1}{2}m_n\dot{u}_n^2+\frac{1}{2}[V_{n-1}(u_n-u_{n-1})+V_n(u_{n+1}-u_n)],\label{e:energy}
\end{equation}
where the interaction potential $V_n(u_{n+1}-u_n)$ is given by
\begin{align}
V_n(u_{n+1}-u_n) = &\frac{2}{5}A_{n,n+1}[\Delta_{n,n+1}-(u_{n+1}-u_n)]_+^{5/2}\nonumber\\
& - A_{n,n+1}\Delta_{n,n+1}^{3/2}(u_{n+1}-u_n)\nonumber\\
& - \frac{2}{5}A_{n,n+1}\Delta_{n,n+1}^{5/2}. \label{e:potential}
\end{align}
	For finite chains of length $N$, we have implemented free boundary conditions thus, the equations of motion for the displacements of the end-beads read:
\begin{align}
m_1\ddot{u}_1&=F_0-A_{1,2}[\Delta_{1,2}-(u_2-u_1)]_+^{3/2}\label{e:leftBC}\\
m_N\ddot{u}_N&=A_{N-1,N}[\Delta_{N-1,N}-(u_N-u_{N-1})]_+^{3/2}-F_0.\label{e:rightBC}
\end{align} 
\begin{table}
\begin{tabular}{l  r r r}
\hline\noalign{\smallskip}
Material & Steel & Aluminum & Brass\\[3pt]
\hline\noalign{\smallskip}
Radius (mm) &4.76 &4.76 &4.76 \\[3pt]
\hline\noalign{\smallskip}
Density ($\text{kg}/\text{m}^3$) &8027.17 &2700 &8500 \\[3pt]
\hline\noalign{\smallskip}
Elastic modulus (Pa) &$1.93\times 10^{11}$ &$7\times 10^{10}$ & $7.6\times 10^{10}$ \\[3pt]
\hline\noalign{\smallskip}
Poisson ratio &0.3 &0.35 &0.414 \\[3pt]
\hline
\end{tabular}
\caption{
	The material properties used in our simulations.
}\label{t:matprops} 
\end{table}
	More specifically, we consider a chain of alternating beads (i.e., a diatomic lattice with the beads having the same radius but being made of different materials) and a chain of a third type of bead (monoatomic) adjacent to the diatomic fraction of the full lattice. 
	Figure \ref{f:setup} shows the schematic of the bead configuration. 
	In the first half of the chain, the alternating beads used consist of steel and aluminum; the second half of the chain is composed entirely of brass beads.
	This specific configuration and these materials were chosen based on their straightforward availability for experimental realizations, and equally (or more) importantly, because of the properties of  the underlying linear spectrum which enable the existence of localized vibrational mode at the interface between the diatomic and the monoatomic part of the chain. 
	We will discuss these linear properties in more detail in a later section. 
	A summary of the material elastic and geometric characteristics used is given in Table \ref{t:matprops}.

\section{Results}
\subsection{Numerical methods}
	The numerical integration of the system was performed using the \verb!ode113! function in Matlab, which is a variable order Adams-Bashforth-Moulton solver, and allows for stringent error control.
	To calculate the numerically exact (up to a prescribed relative tolerance, usually $10^{-10}$) profile of each DB solution, we used a Newton-Raphson solver to calculate the fixed-points of a Poincar\'{e} map with a frequency $f$ and, thus, a period $T=1/f$,
	$$R(\vec{u}(0))\equiv \vec{u}(0)-\vec{u}(T)=0.$$
	Where, we add the additional constraints that $\dot{u}_n=0$ in order to "pin" the solution to a point on the Poincar\'{e} map. 
	The Jacobian for the Newton's method is given by $I-V(T)$, where $V(T)$ is the monodromy matrix, which we obtain by integrating the variational equation
	$$ \frac{d}{dt}V=J(u,\dot{u})V,$$
simultaneously with the equations of motion \eqref{e:motion}, where $J$ is the Jacobian of these nonlinear equations, and $V(0)=I$.
	See \cite{flach2008}, and the references therein for more details.
	The original initial guess for such an interface DB consisted of the eigenmode of the lower cutoff edge of the optical band of the linear spectrum. 
	However, in the case of the bulk DBs, there was no eigenmode that was localized only in the bulk, and we needed to first obtain one solution from a purely diatomic chain, which we then embedded in the full chain as our initial guess.
	We then performed a continuation of the frequency through the linear spectrum band gap \cite{doedel97}. 
	This was done both for the interface and for the bulk DBs.
	It should additionally be noted that, upon convergence of the relevant iteration, the eigenvalues of the monodromy matrix provide us with the Floquet multipliers determining the linear stability of the corresponding solution. 
	The presence of eigenvalues of unit modulus (i.e., of multipliers that reside on the unit circle) is a signature of linear stability of the DB modes, while the existence of eigenvalues with modulus greater than one is a signature of (linear) instability, as discussed also in more detail below.

\begin{figure}
\includegraphics[width=.45\textwidth]{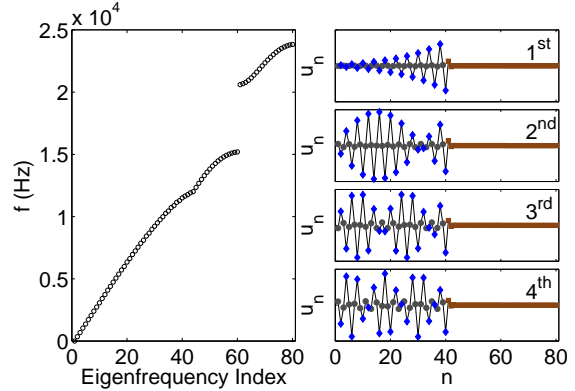}
\caption{
	(Color online)
	In the left panel, the eigenfrequencies of the configuration shown in Fig. \ref{f:setup} are shown for 80 beads.
	The right four panels show the first, second, third, and fourth optical eigenmodes. 
	For clarity, steel beads are shown as gray \emph{circles}, aluminum beads are shown as blue \emph{diamonds}, and brass beads are shown as brown \emph{squares}.
}\label{sf:linear_spectrum}
\end{figure}

\subsection{Underlying linear spectrum}\label{s:linear}
The eigenfrequencies of the configuration of Fig. \ref{f:setup}, are shown in Fig. \ref{sf:linear_spectrum}. 
	These are computed by eigenanalysis of the linearized Eqs. (\ref{e:motion}), (\ref{e:leftBC}), and (\ref{e:rightBC}).
	The linear spectrum of the diatomic part of the chain contains two branches (\emph{acoustic} and \emph{optical}) and possesses a gap between the upper cutoff frequency $\omega^{d}_{1}=\sqrt{2K_{SA}/m_{S}}$ of the acoustic branch and the lower cutoff frequency $\omega^{d}_{2}=\sqrt{2K_{SA}/m_{A}}$ of the optical branch, where $m_{S}$ ($m_{A}$) is the mass of the steel (aluminum) beads. 
	The linear stiffness of the steel-aluminum contact is $K_{SA}=\frac{3}{2}A_{SA}^{2/3}F_0^{1/3}$. 
	The upper cutoff frequency of the optical band is located at $\omega^{d}_{3}=\sqrt{2K_{SA}(1/m_{A}+1/m_{S})}$.
	The linear spectrum of the monoatomic part contains only an acoustic branch with frequencies from $0$ to $\omega^{m}_{1}=\sqrt{4K_{BB}/m_{B}}$, where $m_{B}$ is the mass of the brass beads and $K_{BB}=\frac{3}{2}A_{BB}^{2/3}F_0^{1/3}$ the linear stiffness of the brass-brass contact.

	For the material properties mentioned above and for a precompressive force of $F_{0}=22.4$ N, the cutoff frequencies obtain the following values: $f^{m}_{1}=15.21$kHz, $f^{d}_{1}=11.96$kHz, $f^{d}_{2}=20.6$kHz, and $f^{d}_{3}=23.8$kHz. 
	Since the upper edge of the acoustic band of the monoatomic part of the chain is less than the lower edge of the optical band for the diatomic part of the chain, we obtain a gap in the spectrum of the union of the two parts of the chain, between $f^{m}_{1}=15.21$kHz and $f^{d}_{2}=20.6$kHz. 
	The materials we used were chosen in such a way that the acoustic band of the monoatomic part of the chain did not completely overshadow the band gap in the diatomic part of the chain. 
	We can widen or narrow the gap based on the materials chosen and on the magnitude of the precompressive force.

	The eigenmode from the bottom edge of the optical band was used as an initial guess in our Newton's method solver. 
	An example guess, for 80 beads, is shown in the top of the right panels of Fig. \ref{sf:linear_spectrum}. 
	Later, the second and third optical eigenmodes were used as an initial guess in continuations through the gap. 
	These eigenmodes are shown in Fig. \ref{sf:linear_spectrum} as well.

\begin{figure}
\includegraphics[width=.45\textwidth]{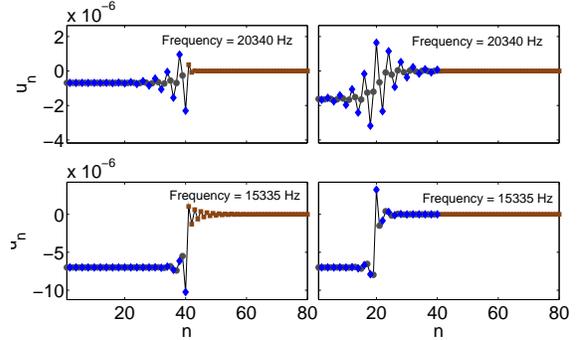}
\caption{
	(Color online)
	Examples of four solutions. 
	The left two are located at the interface, the right two are in the bulk; the top two are closer to the optical band edge, while the bottom two are deeper within the spectral gap, and thus exhibit a more pronounced localization.
	For clarity, steel beads are shown as gray circles, aluminum beads are shown as blue diamonds, and brass beads are shown as brown squares.
}\label{f:breather_profiles}
\end{figure}

	As one can see, the eigenmode from the bottom edge of the optical band is localized right on the interface between the two halves of the chain. 
	This localization is caused by the use of free boundary condition (BC) on the left and the interface between an aluminum and brass bead. 
	In an infinite diatomic chain, the lower optical mode has a profile in which the heavy particles (steel in our case) are at rest while the light particles (aluminum) move in an alternating pattern with the same amplitude. 
	In a semi-infinite diatomic chain, with fixed left BC, [$u(0)=0$] when the first particle is light, the form of this mode is exactly the same as that of the inifinite chain since the ``wall'', particle 0, is a heavy particle at rest as it should be.
	However, using a heavy particle at the first site, or alternatively, a free left BC [$u(0)=u(1)$], leads to a lower optical mode in which the light masses move again in an alternate pattern but now with a decreasing amplitude as we approach the left part of the chain. 
	In our configuration, the diatomic part of the chain has a left free BC and ends with a light particle at the interface. 
	The latter acts more like a wall, rather than a free end; the brass bead is relatively heavy, with a mass close to that of a steel bead, and during the excitation of the lower optical mode, the displacement of the brass bead is very small around the equilibrium since this frequency belongs to the gap of the underlying monoatomic spectrum of the brass. 
        On the other hand, according to the above explanation, an interface between a steel and brass bead leads to a mode with substantially decayed amplitude at the interface. 
        Thus, a configuration with that kind of interface cannot lead to an interface breather that bifurcates from the linear limit of the lower edge of the optical branch.

\begin{figure}
\includegraphics[width=.45\textwidth]{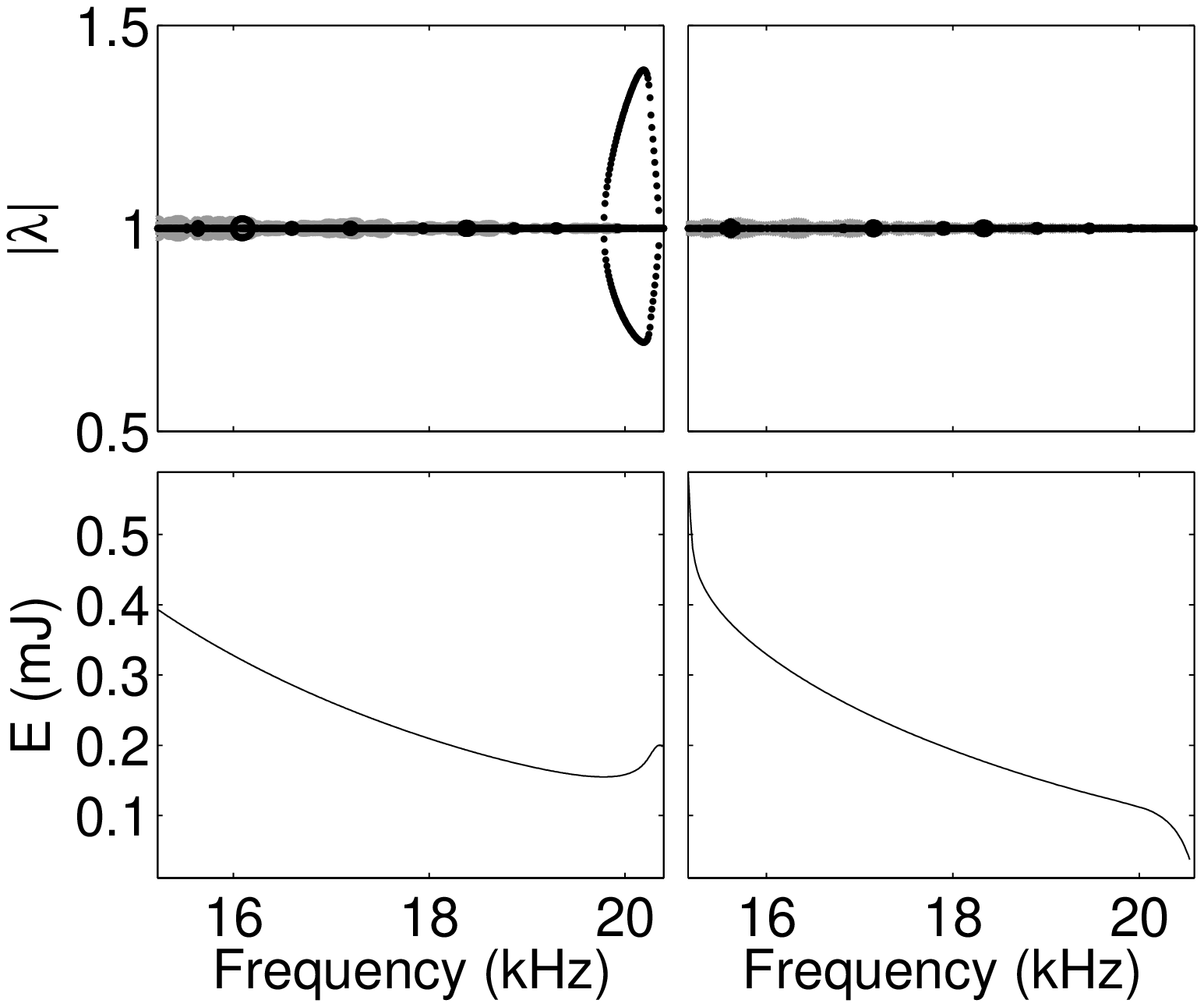}
\caption{
	\emph{Top:} the magnitudes of the Floquet multipliers for the bulk (\emph{left}) and interface (\emph{right}) breathers are shown as a function of frequency. 
	In both cases, the \emph{real} multipliers are shown in black, while the \emph{complex} multipliers are shown in gray. 
	\emph{Bottom:} the energies of the bulk (\emph{left}) and interface (\emph{right}) breathers are shown as a function of frequency. 
	Note that the vertical scales are the same in each row, and the horizontal scales are the same in each column.
}\label{f:ababcc_floq_energy}
\end{figure}

\subsection{Existence and stability of interface discrete breathers}\label{s:breathers}
	Using the eigenmode from the lower edge of the optical band as an initial guess, we performed a continuation as a function of the frequency through the linear spectral gap. 
	This resulted in the identification of the DBs that were located at the interface between the diatomic and monoatomic chains, to which we will refer as interface DBs. 
	In addition, for comparison, we used the corresponding linear eigenmode of a purely diatomic chain to solve for DBs in the bulk of the diatomic part of the chain (i.e. away from the interface and boundaries). 
	We will refer to the latter class of solutions as bulk DBs. 
	Some examples of typical obtained DB profiles are shown in Fig. \ref{f:breather_profiles}. 
	The further into the middle of the gap the frequency of the breather is located, the more localized the relevant structure becomes (a feature shared by the interface and bulk waveforms). 
	Approaching the acoustic branch, both interface and bulk waveforms delocalize and obtain non-zero oscillating tails due to resonance with the upper acoustic cutoff mode of the monoatomic brass part. 
	More details about this kind of delocalization for the case of the bulk breathers in a diatomic chain can be found in \cite{Theo2010}.

	On the other hand, the linear stability analysis illustrated one major difference between the bulk breathers and their interface analogs. 
	For both types of DBs, the total energy of each breather solution ($E=\sum_{n=1}^{N}e_{n}$) was computed as a function of the corresponding frequency (during the frequency continuation). 
	In the plot of energy vs. frequency for the bulk case, there is a {\it change in the monotonicity}. 
	This has been found to correspond to a change in the stability of the solutions (shown in Fig. \ref{f:ababcc_floq_energy}) (notice that this type of result suggests more broadly that a theorem of this type could be mathematically proved, by analogy to the famous Vakhitov-Kolokolov criterion for stability of solutions in nonlinear Schr{\"o}dinger type models \cite{VK}). 
	More specifically, when the energy of the solution is decreasing with respect to its frequency, the corresponding breather is either linearly stable (e.g., very close to the optical band edge) or only oscillatorily unstable. 
	The latter oscillatory instabilities are due to quartets of Floquet multipliers, with non-zero imaginary parts, that depart from the unit circle. 
	Based on our observations, the magnitude of oscillatory instabilities decreases as the length of the chain increases, so these are essentially weak instabilities that only manifest themselves over very long time scales (much longer than the potential experimental lifetime of the breather \cite{boechler2009}). 
	One important note about these bulk DBs is that, unlike the case of a purely diatomic chain, they do not bifurcate from the optical band, and because of this, we could not perform the continuation all the way to the top of the gap (once again, see \cite{Theo2010} for more details about the purely diatomic case).
	On the other hand, when the energy of the solution is increasing with respect to the frequency, the corresponding breather shows a definitive signature of a {\it real instability}, corresponding to a pair of Floquet multipliers bifurcating off of $(1,0)$ and on to the real axis. 
	See Fig. \ref{f:ababcc_floq_energy} for a more detailed look at the Floquet multipliers. 
\begin{figure}[t]
\includegraphics[width=.45\textwidth]{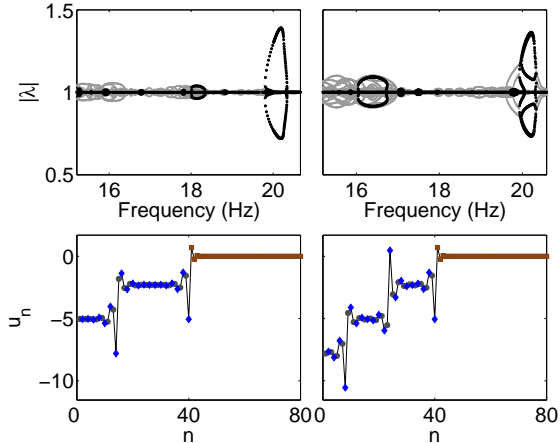}
\caption{
	(Color online)
	\emph{Top:} the magnitude of the Floquet multipliers for the double (\emph{left}) and triple (\emph{right}) breathers is shown as a function of frequency. 
	In both cases, the \emph{real} multipliers are shown in black, while the \emph{complex} multipliers are shown in gray. 
	\emph{Bottom:} example profiles of the double (\emph{left}) and triple (\emph{right}) breathers for a frequency of 18.3 kHz.
	For clarity, steel beads are shown as gray circles, aluminum beads are shown as blue diamonds, and brass beads are shown as brown squares.
}\label{f:ababcc_doub_trip_comparison}
\end{figure}
	Remarkably, in the plot of energy vs. frequency for the interface breathers (shown in Fig. \ref{f:ababcc_floq_energy}), there is {\it no such change in monotonicity}, leading to the conclusion that throughout the parametric continuation, the interface breathers are either stable or merely subject to weak oscillatory instabilities. 
	This suggests that the interface is acting as a stabilizer for the breathers and may be useful for experimental purposes of ensuring the robustness/observability of such structures especially in the vicinity of interfaces. 
	It is even more remarkable that similar (yet reverse in their nature) stability modification conclusions have been reported in the literature of nonlinear Schr{\"o}dinger type systems, see e.g., \cite{makris2005}, and \cite{suntsov2006}, where it was found that surface breathers may be less robust (and subject to an interval of instability) than their bulk counterparts. 
	However, it should be noted that the latter findings are for lattices that even in their monoatomic realization possess localized solutions of this type (whereas for the granular chains considered here, a diatomic setting is necessary for the existence of bulk breathers within the linear spectral gap \cite{Theo2009,boechler2009}).

\subsection{Continuation of second and third optical eigenmode}
	The first optical eigenmode was not the only eigenmode that may lead to  localization upon continuation within the spectral gap. 
	Figure \ref{sf:linear_spectrum} shows that the second and third eigenmodes also led, upon corresponding continuations, to states with two or three, respectively, breather structures, with one localized at the interface in each case. 
	This trend can be extended to higher eigenmodes which similarly lead to corresponding multi-breather structures within the spectral gap.
	We performed a continuation through the gap using the second, and then the third, optical eigenmode as an initial guess (once again, with 80 beads). 
	Figure \ref{f:ababcc_doub_trip_comparison} shows a sample profile from each of these continuations (as well as their parametric stability properties). 
	The resulting solutions contained two, and three, breathers, respectively, which can be clearly discerned to develop near the regions of maximal displacement in the respective eigenmodes. 
	Once again, the linear stability of each solution was analyzed using the Floquet multipliers. 
	A detailed view of the obtained multipliers is shown for the double and triple breathers in Fig. \ref{f:ababcc_doub_trip_comparison}. 
	These double and triple breathers have intervals of instability similar to the bulk breathers, which implies that it is the bulk breathers ``within the structure'' that become unstable, while the localization at the interface may be expected to
persist.
	This is confirmed by dynamical simulations of these breathers, in which we make the following observations.
	When we perturb an unstable double, or triple breather, the unstable bulk part radiates some energy and decays to one of the stable bulk breathers (with a lower frequency), while the interface breather remains unchanged.
	Then the two (or three) breathers oscillate out of sync, until oscillatory instabilities disperse the breathers: usually the bulk breathers first, followed by the interface breather.
\begin{figure}
\includegraphics[width=.45\textwidth]{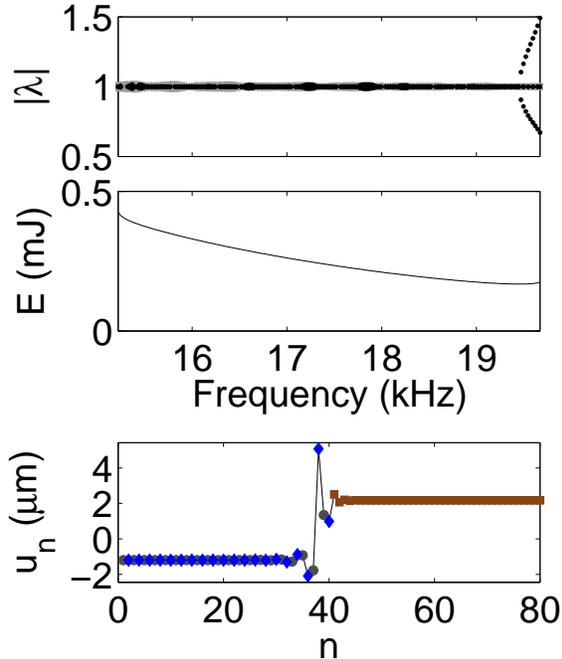}
\caption{
	(Color online)
	Results for a continuation of a DB located one unit cell from the interface.
	\emph{Top:} the magnitude of the Floquet multipliers for each DB is shown as a function of frequency. The purely real multipliers are highlighted in black.
	\emph{Middle:} the energy of each DB is shown as a function of frequency.
	\emph{Bottom:} a sample DB solution from this family is shown.
	Note that for visual clarity, brass is shown in brown (squares), steel is shown in gray (circles), and aluminum is shown in blue (diamonds).
}\label{f:ababcc_family2_floq_energy_prof}
\end{figure}
\subsection{Another family of discrete breathers}
	Although in the above considerations, we examined the different families of breathers arising from the linear limit, an interesting question concerns whether additional families of surface breathers resembling the ones above may exist, shifted away from the interface; such families may, in fact, cease to exist before the linear limit is reached since they do not degenerate into linear modes. 
	For example, as we will see in the dynamical evolution simulations below, under certain conditions (of interaction of a traveling compression pulse with an interface breather), a different type of surface mode may be excited.
	In this mode, the maximal displacement site is shifted away from the interface by one unit cell (two lattice sites).
	Using an appropriate initial guess of a breather with an appropriately shifted center, we performed a continuation throughout the frequency gap, and obtained such a family of breathers which were localized one unit cell away from the interface.
	The bottom panel in Fig. \ref{f:ababcc_family2_floq_energy_prof} shows a sample profile from this continuation, while the top two panels show the magnitude of the Floquet multipliers, and the energy as a function of the DB frequency.
	It is interesting to note that this family of breathers does not bifurcate from the optical band, as the other family of interface DBs does.
	In fact, as the frequency of these DBs approaches the optical band, they develop a real instability (cf. also the associated change of monotonicity in the energy-frequency plot in Fig. \ref{f:ababcc_family2_floq_energy_prof}). 
These real instabilities increase rapidly until the family undergoes a
saddle-center bifurcation, and ceases to exist.
\begin{figure}[t]
\includegraphics[width=.45\textwidth]{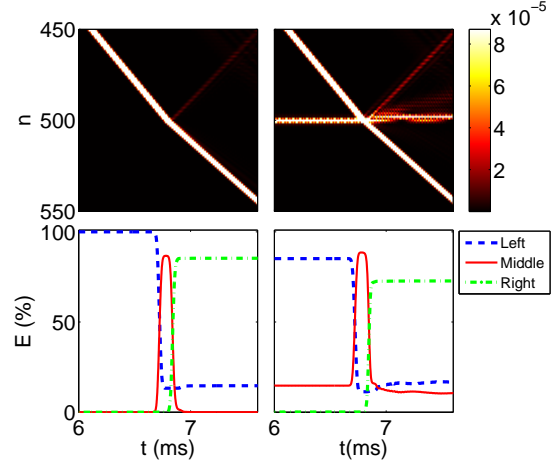}
\caption{(Color online)
	\emph{Top left:} a spacetime plot of the local energy density for a compression pulse passing through the interface. 
	There is a barely visible reflection amounting to about 2\% of the total energy. 
	The \emph{lower left} panel shows a different view of the same simulation.
	Here, each line represents the total energy in a certain part of the chain. 
	The dashed line (blue) represents the first half of the chain, the solid line (red) represents the interface region, and the dash-dotted line (green) represents the second half of the chain. 
	There is no significant amount of energy trapped in the interface. 
	The \emph{right} two panels show a similar picture, except that in this case, there is an interface breather for the pulse to interact with. 
	The pulse causes the breather to shift by 1 cell. 
	Note that the time-scale is the same for the top and bottom panels, and the color scale for the top panels is truncated at 1/20 the maximum, in order for the reflections to be visible.
}\label{f:ababcc_pulse_interface}
\end{figure}

\begin{figure}[t]
\includegraphics[width=.45\textwidth]{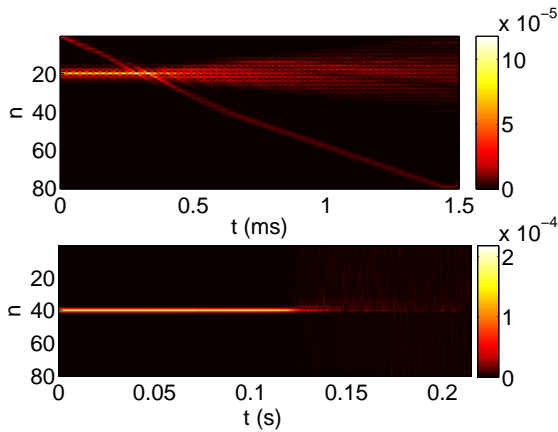}
\caption{(Color online)
	\emph{Top:} the destruction of a bulk breather due to interaction with a compression pulse.
	\emph{Bottom:} the destruction of an interface breather due to the oscillatory instabilities.
}\label{f:breather_decay}
\end{figure}

\subsection{Discrete breather and nonlinear interface dynamics}
	We now turn to direct numerical simulations in order to examine the nonlinear dynamics of the chain (and, in particular, of the interface), as well as to assess the role of the discrete interface breathers therein. 

	Our first set of numerical experiments involved scattering events. 
	The aim of these was to investigate the behavior of a compression pulse as it interacted with the interface, in the absence, as well as in the presence of an excited breather. 
	We excited the first bead with a striker to send a compression pulse through the chain.
	Many of these simulations were conducted over a wide range of velocities for the striker, and Fig. \ref{f:ababcc_pulse_interface} shows typical examples within this class of numerical simulations. 
	In this particular simulation, the velocity of the striker was about 0.94 m/s, and produced a solitary wave with an amplitude of about $2.5\times 10^{-5}$ m and a maximum dynamical force of approximately 138 N, but similar results were obtained for lower initial striker velocities.
	Note that in this figure, the space-time contour plots of the local energy density (defined above), rather than the displacements, are  shown.  
	The leftmost plots in Fig. \ref{f:ababcc_pulse_interface} show a compression pulse colliding with the interface; although we observe some mild reflection (about 2\%; see also below) at the interface, it is generally found that the pulse does {\it not} excite a localized mode at the interface.
	The rightmost plots in Fig. \ref{f:ababcc_pulse_interface} show a compression pulse colliding with a pre-existing breather at the interface.
	 Surprisingly, there does not appear to be a significant difference in the behavior of the pulse in this second case, but it does illustrate that under certain circumstances, the breather actually shifts away from the interface upon collision with the pulse. 
	 The dynamics here clearly appears to lead to a member of the family of surface discrete breathers studied in the previous section. 
	After further tests, we conclude that whether or not a breather shifts away from the interface appears to depend on its linear stability. 
	It seems to be the case that a compression pulse of the right magnitude will shift an interface breather away from the interface only if the breather has a pair of Floquet multipliers that split off the unit circle at the $(-1,0)$ point of the unit circle. 
	I.e., such subharmonic instabilities can lead to potential mobility of the interface breathers under the action of suitably strong perturbations. 
	In order to measure the transmission of energy past the interface, we measured the change in energy in each part of the chain (the first diatomic half, the interface, and the monoatomic second half), and divided by the total  energy.
	 The bottom two panels of Fig. \ref{f:ababcc_pulse_interface} show a typical example of how the energy propagated through the chain (in this case, for a chain of 1000 beads). 
	 The left hand part is the total energy in beads 1-497, the interface part is the total energy in beads 498-502, and the right hand part is the total energy in beads 503-1000. 
	The reflection of the solitary wave from the interface is observed as the small dip in the dashed line after the pulse passes through the interface and only amounts to about 2\% of the total energy.
	Notice that although the reflection is fairly weak, the fraction of energy remaining on the left part of the domain is more substantial; this is due to the compression pulse formation process from the striker, which has led to residual ``radiation'' of phonons in its wake (not shown here, since for clarity we only show the part of the domain in the vicinity of the interface).
		
\begin{figure}
\includegraphics[width=.45\textwidth]{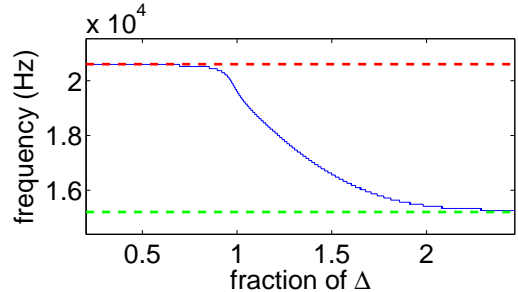}
\caption{(Color online)
	The frequency of the induced breather (or linear mode) when the middle site (to the left of the interface) was excited with a negative displacement is shown as a function of the magnitude of the initial excitation.
	Recall that $\Delta$ is the amount of overlap between the beads at equilibrium.
	The dashed lines represent the acoustic and optical band edges.
}\label{f:excited_middle_frequencies}
\end{figure}

	 On the other hand, we note in passing that, in the case of the unstable bulk breathers (through a real pair of Floquet multipliers), their interaction with a compression pulse has been observed to lead to the ultimate destruction (i.e., dispersion) of the breather mode, which is shown in Fig. \ref{f:breather_decay}.
	 Lastly, even in the absence of the solitary wave, the interface breathers will eventually decay, due to the oscillatory instabilities, as shown in Fig. \ref{f:breather_decay}. 
	 This is a similar phenomenon to what is observed also for bulk breathers when they are subject to the same type of instabilities \cite{Theo2010}.

	Another set of performed numerical experiments concerned the realization of the minimum initial excitation needed to produce an interface breather with a frequency in the spectral gap. 
	From the simulations above, it was clear that compression pulses did not excite such a breather, regardless of which end they started from. 
	This is similar to what is shown in the bottom left panel of Fig. \ref{f:ababcc_pulse_interface} where it shows the interface region has a brief spike as the pulse passes through the interface, but then returns to zero.
	Exciting the middle site with a positive displacement was also unsuccessful.
	The positive displacement excites the monoatomic part of the chain, namely acoustic modes up to 15.21kHz. 
	But this part of the chain is dispersive and the nonlinearity can not support localized vibrations. 
	Thus, in all cases of this type, the energy was dispersed into linear waves and nonlinear solitary waves, and no substantial fraction thereof was found to be trapped at the interface.

\begin{figure}
\includegraphics[width=.45\textwidth]{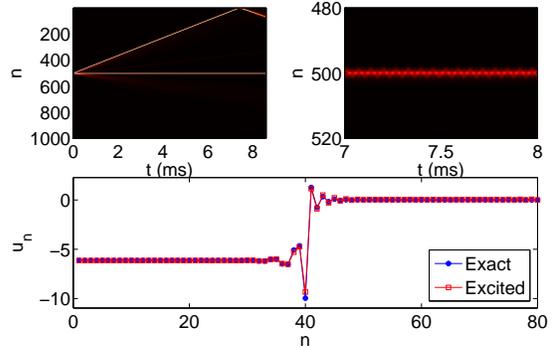}
\caption{(Color online)
	\emph{Top left}: The space-time graph showing the energy density for a chain of 1000 beads in which the 500th bead was excited to about $1.5 \Delta$. 
	After some initial radiation, a DB is excited at the interface. 
	\emph{Top right:} A zoom of the top left panel, showing that there is indeed a DB located at the interface. 
	\emph{Bottom:} the profile of the excited breather is shown in conjunction with the exact breather for the corresponding frequency.
}\label{f:excited_middle}
\end{figure}

	On the other hand, when the middle site was excited with a negative displacement of sufficient amplitude, we observed the formation of discrete interface breathers.
	The larger the amplitude of the initial excitation, the lower the frequency into the gap the resulting breather is found to be (Fig. \ref{f:excited_middle_frequencies} shows the frequency of the resulting breather as a function of the initial excitation). 
	The top panels of Fig. \ref{f:excited_middle} show the results of a simulation in which the central site was excited to about $1.5\Delta$. 
	It can be observed that, after some initial solitary waves are released, a robust interface breather is formed almost immediately. 
	The resulting breather, which, in this case, has a profile shown in Fig. \ref{f:excited_middle}, has a frequency of about $16.193$kHz, and matches almost perfectly the exact breather of that frequency identified through the Newton's method.
	These observations may be of use in trying to produce these modes experimentally.

\section{Conclusions and Future Challenges}
	In the present work, we considered the possibility of formation of discrete breathers at the interface between a diatomic chain and a monoatomic chain (of same radii beads but made of different materials). 
	We chose this particular configuration due to its underlying linear properties and elucidated how (and under what conditions) these give rise to a linear spectral gap whose optical band cutoff mode enables a continuation towards a discrete breather state. 
	The continuation of such modes was performed throughout the spectral gap and the apparent robustness of these structures was elucidated. 
	It was, in fact, illustrated that these modes are arguably more robust than their bulk counterparts possessing no parametric intervals of non-unit Floquet multipliers along the real axis (as their bulk counterparts do), but being only subject to weak oscillatory instabilities for finite-size chains.
	This should imply that similarly to what was shown for their bulk counterparts in~\cite{boechler2009}, such modes should be possible to experimentally produce in granular chains.
	In this general direction of interfaces, the work of ~\cite{Daraio3} has experimentally considered the propagation of waves at the interface of two different granular chains.
	More recently, in the same spirit, the work of~\cite{khatri2009} has focused on the dynamics of compression pulses at the interface of an uncompressed monoatomic chain with a rod. 
	In our setting, in addition to illustrating the existence and stability of breather interface modes, we also examined how they behave dynamically and how they interact with compression pulses. 
	While the latter are unable to excite such interface breathers, we have illustrated under what conditions (and negative displacements of the interface bead) such modes may arise. 
	Furthermore, we have shown that their interaction with compression pulses may lead to their shift from the interface. 
	On the other hand, it is interesting to mention that we have also revealed multi-breather states containing both breathers located at the interface, as well as (one or more) bulk ones. 
	These stem from higher order linear modes of the spectral problem involving the interface.
	Lastly, we have shown the existence of a family of interface DBs, which does not bifurcate from the optical band, as do the other families of interface DBs.

	Our results suggest a number of interesting directions for future explorations.
	On the one hand, they raise the question of appropriate quantitative diagnostics for the examination of interaction of pulses (or breathers) with interfaces. 
	On the other hand, a natural topic that emerges concerns the conditions under which a propagating excitation (e.g. a nearly compact pulse) may produce energy trapping at an interface similar to the ones used herein. 
	Finally, it would be interesting to examine similar phenomena (and interfacial dynamics) in other classes of models, such as a binary interfacing with a unary waveguide array modeled by the discrete nonlinear Schr{\"o}dinger equation to examine the potential generality of the conclusions obtained herein [in particular, ones about the enhanced stability of the surface modes].
	Such investigations are presently underway and will be reported in future publications.

\acknowledgments This work has been supported from "A.S. Onassis" Foundation, RZG 003/2010-2011 (GT and PGK). PGK gratefully acknowledges support from the National Science Foundation through grants
NSF-DMS-0349023 (CAREER), NSF-DMS-0806762 and the Alexander von Humboldt Foundation.
The authors acknowledge numerous enlightening discussions with Prof. Chiara Daraio, and are thankful to Prof. Daraio for stimulating their interest in this research topic and for providing the preprint of~\cite{khatri2009}  prior to its publication.
	A number of discussions at an early stage of this project with Prof. Mason .A. Porter are also acknowledged.


\end{document}